\documentclass[12pt]{article}
\usepackage{geometry,amsmath,amssymb,epsfig}
\newcommand{\Omicron}{\mathrm{O}}
\newcommand{\emdash}{---}
\newcommand{\mathe}{\mathrm{e}}
\newcommand{\tmem}[1]{{\em #1\/}}
\newcommand{\tmop}[1]{\ensuremath{\operatorname{#1}}}
\begin{document}

\title{Anomalous discrete chiral symmetry in the Gross-Neveu model and
loop gas\\
simulations}
\author{
Oliver B\"ar, Willi Rath and Ulli Wolff\thanks{
e-mail: uwolff@physik.hu-berlin.de} \\
Institut f\"ur Physik, Humboldt Universit\"at\\ 
Newtonstr. 15 \\ 
12489 Berlin, Germany
}
\date{}
\maketitle

\begin{abstract}
  We investigate the discrete chiral transformation of a Majorana fermion on a
  torus. Depending on the boundary conditions the integration measure can
  change sign. Taking this anomalous behavior into account we define a chiral
  order parameter as a ratio of partition functions with differing boundary
  conditions. Then the lattice realization of the Gross-Neveu model with
  Wilson fermions is simulated
  using the recent `worm' technique on the loop gas or all-order hopping
  representation of the fermions. An algorithm is formulated that includes the
  Gross-Neveu interaction for $N$ fermion species. The critical line $m_c (g)$
  is constructed for a range of couplings at $N = 6$ and for $N = 2$, the
  Thirring model, as examples.
\end{abstract}
\begin{flushright} HU-EP-09/24 \end{flushright}
\begin{flushright} SFB/CCP-09-44 \end{flushright}
\thispagestyle{empty}
\newpage

\section{Introduction}

Many textbooks on quantum field theory start with scalar fields and discuss,
after preparing the arena with non-interacting free fields, the quartic
self-coupling as the simplest conceivable interaction. At this stage
renormalization or, on the lattice, the continuum limit has to be discussed in
a non-trivial way. A mild generalization leads to $N$ real scalar fields
self-coupled by a unique quartic O($N)$ invariant term. As is well known the
kinetic term dictates the field dimension and implies a mass dimension $4 - D$
for the coupling constant. Thus the self-coupled scalar theory is
renormalizable in $D = 4.$ In fact, $D = N = 4$ is the standard model in the
limit where all fields except the Higgs are stripped away. The perturbative
renormalization group $\beta$-function is positive for small coupling and the
models are hence perturbatively trivial. Simulations with simple lattice
discretizations suggest that this features is also true nonperturbatively.

An analogous fermionic model can be written in terms of $N$ `real' Majorana
fermions. Now, because there is only one derivative in the fermionic kinetic
term, a quartic coupling has dimension $2 - D$. The renormalizable theory in
$D = 2$ has in fact again a unique O($N)$ invariant interaction. It has been
proposed and considered in the limit of $N \rightarrow \infty$ a long time ago
{\cite{Gross:1974jv}} and is now known as the Gross-Neveu (GN) model. For the
sign of the quartic interaction that has been discussed to be the physically
meaningful one in {\cite{Gross:1974jv}} the GN model is even asymptotically
free. This feature that is shared by QCD was one of the main reasons for 
Gross and Neveu to
study the model as an analog toy-model. The analogy goes even further as there
also is a chiral symmetry in the GN model. Although it is only discrete,
this symmetry involves the two-dimensional analog of $\gamma_5$ and is only
present for massless fermions, which are thus distinguished by an enhanced
symmetry. Gross and Neveu have shown that the chiral symmetry breaks spontaneously (at
infinite $N)$ and that mass generation and `dimensional transmutation' take
place. This beautiful parameter-free theory is then an analog of QCD in the
chiral limit.

There have been some early efforts {\cite{Cohen:1983nr}} to study this model
beyond perturbation theory by Monte Carlo simulation. Attempts employing
today's standard techniques used for QCD {\cite{Duane:1987de}} were made in
{\cite{korzec:2006hy}} and {\cite{TomPhD}}. However in {\cite{Wolff:2007ip}}
and {\cite{Wolff:2008xa}} we have proposed techniques to simulate the
fermionic Grassmann integral in a more direct and efficient way using its loop
gas representation {\cite{Gattringer:1998cd}}, see also {\cite{Wenger:2008tq}}
for a related effort. In this formulation, using Wilson fermions, the
simulations resemble those for simple spin models with efficient algorithms.
No nearly singular inversions and no nonlocal effective bosonic actions occur.
While in more than two dimensions a sign problem still prevents the
application of this technique {\cite{Wolff:2008xa}} it is well suited for the
GN model as we demonstrate below.

Like in QCD, the important chiral symmetry is violated by the Wilson
discretization and an additive mass renormalization, i.e. a nontrivial
definition of the critical point, is required. In this article we discuss,
that, depending on the boundary conditions on a finite torus, an extra minus
sign from the non-invariance of the Majorana fermion measure has to be taken
into account in symmetry relations. We consider this as (a primitive form of)
an anomaly. One such symmetry relation allows us to define a finite volume
chiral order parameter. This way of defining the chiral point in parameter
space is particularly convenient in the loop gas representation, where it
corresponds to a well measurable topological variable.

After a definition of the model in section~2 the anomaly is discussed for the
continuum theory in section~3 and connected with the fluctuating boundary
conditions in section~4. In section~5 the loop gas representation and simulation on
the lattice are described and numerical results are reported in section~6. Finally we
end in section~7 with conclusions and an outlook.

\section{O($N$) invariant Gross Neveu model}

The O($N$) invariant GN model {\cite{Gross:1974jv}} is best written in terms
of $N$ self-coupled Majorana fermions $\xi_{\alpha, a} (x)$ with the Euclidean
action
\begin{equation}
  S = \int d^2 x \frac{1}{2}  \overline{\xi} (\gamma_{\mu} \partial_{\mu} + m)
  \xi - \frac{g^2}{8} ( \overline{\xi} \xi)^2, \label{SGN}
\end{equation}
where the contracted index $a = 1, \ldots, N$ refers to the global `flavor'
symmetry and $\alpha = 1, 2$ is spin. For a Majorana fermion $\overline{\xi}$
is not independent but just an abbreviation for
\begin{equation}
  \text{$\overline{\xi}$=} \xi^{\top} \mathcal{C}
\end{equation}
with the 2-dimensional Dirac matrices $\gamma_0, \gamma_1$ and the
charge conjugation matrix $\mathcal{C}$ satisfying
\begin{equation}
  \gamma_{\mu}^{\top} = -\mathcal{C} \gamma_{\mu} \mathcal{C}^{- 1},
  \hspace{1em} \mathcal{C}= -\mathcal{C}^{\top}, \hspace{1em} \{\gamma_{\mu},
  \gamma_{\nu}\} = 2 \delta_{\mu \nu}, \hspace{1em} (\gamma_{\mu})^{\dag} =
  \gamma_{\mu} .
\end{equation}
In two dimensions the model is renormalizable in the strict sense, no other
four-fermion interaction is compatible with the global O($N$). Moreover, the
models with $N \geqslant 3$ are perturbatively asymptotically free. The
Thirring model $N = 2$ has no divergent coupling renormalization. At vanishing
fermion mass $m = 0$, there is in addition a discrete chiral invariance
\begin{equation}
  \xi \rightarrow \gamma_5 \xi, \hspace{1em} \gamma_5 = i \gamma_0 \gamma_1,
  \hspace{1em} [\Rightarrow \overline{\xi} \rightarrow - \overline{\xi}
  \gamma_5] . \label{chisym}
\end{equation}
A standard step in treating the GN model, both in the continuum and on the
lattice, is to factorize the interaction into bilinears at the expense of
introducing a scalar $\sigma (x)$. Foreclosing for a moment the lattice
version, where $d^2 x$ becomes the cell volume $a^2$, we write in the
Boltzmann factor at each $x$
\begin{equation}
  \mathe^{\frac{a^2 g^2}{8} ( \overline{\xi} \xi)^2} = \int_{-
  \infty}^{\infty} \frac{a d \sigma}{\sqrt{2 \pi}} \mathe^{- \frac{a^2}{2}
  \sigma^2 - \frac{a^2}{2} g \sigma \overline{\xi} \xi} .
\end{equation}
This may be formally proven by shifting $\sigma$, but perhaps more
convincingly, we state that both sides agree as polynomials in the nilpotent
Grassmann bilinear [$(\overline{\xi} \xi)^{(N+1)}=0$].

After the introduction of $\sigma (x)$ we are led to the fermionic partition
function
\begin{equation}
  Z (m + g \sigma ; \varepsilon) = \int D_{\varepsilon} \xi \mathe^{-
  \frac{1}{2} \int d^2 x \overline{\xi} (\gamma_{\mu} \partial_{\mu} + m + g
  \sigma) \xi} = \tmop{Pf} [\mathcal{C}(\gamma_{\mu} \partial_{\mu} + m + g
  \sigma)] \label{Pf}
\end{equation}
as the basic fermionic building block. In this formula $\xi$ represents only
one flavor. We imagine this fermionic functional integral here formally on a
finite torus of extent $L_0 \times L_1$ in the continuum. The two component
(`bits') object $\varepsilon$ with $\varepsilon_{\mu} \in \{0, 1\}$ labels
four possible (anti)periodic boundary conditions
\begin{equation}
  \xi (x \pm \hat{\mu} L_{\mu}) = (- 1)^{\varepsilon_{\mu}} \xi (x) .
\end{equation}
The Pfaffian Pf is the counterpart of the determinant associated with Dirac
fermions. Note that Pf is only defined for antisymmetric matrices and that the
operator in (\ref{Pf}) [absorbing $m$ and $g$ now into $\sigma$]
\begin{equation}
  A (\sigma) =\mathcal{C}(\gamma_{\mu} \partial_{\mu} + \sigma) = - A^{\top}
  (\sigma) \label{Adef}
\end{equation}
indeed has this property for all four choices of $\varepsilon$. One relation
for Pfaffians is that they square to determinants,
\begin{equation}
  (\tmop{Pf} [A (\sigma)])^2 = \tmop{Det} [A (\sigma)].
\end{equation}
Note that the sign of $\tmop{Pf}$, which is our focus in the next
section, is not determined by the determinant.

\section{Symmetry of the Pfaffian in a finite volume}

By a suitable choice of Dirac matrices $A$ can be taken to be real. Hence its
eigenvalues are imaginary and come in pairs
\begin{equation}
  A (\sigma) f = i \lambda f, \hspace{1em} A (\sigma) f^{\ast} = - i \lambda
  f^{\ast} .
\end{equation}
In addition, by using $\gamma_5$, it is trivial to demonstrate that $A (-
\sigma)$ possesses precisely the {\tmem{same pairs}} of eigenvalues. By
arguments analogous to those used for Weyl fermions{\footnote{We admit here
that our arguments about infinite objects disregarding renormalization are at
a highly formal level. However, this will be substantiated on the lattice in
our case.}} in connection with the global anomaly {\cite{Witten:1982fp}} a
functional definition of the Pfaffian as a root of the determinant has to be
smooth under deformations of $\sigma$. For some reference field $\sigma$ where
there are no vanishing eigenvalues we define the Pfaffian by multiplying all
positive eigenvalues $\lambda$ only. This represents a certain definition of
the (irrelevant) over-all phase. A possible simple reference field would be
$\sigma (x) \equiv m > 0$, the free massive Majorana fermion. Then the
Pfaffian for other field configurations including its sign follows from
smoothness.

Again as in {\cite{Witten:1982fp}} this requires us to follow the flow of
individual eigenvalues as functionals of $\sigma$. As a pair of them crosses
zero the associated eigenvectors allow us to follow the members of the pair
separately and defines a unique continuation of the phase of the Pfaffian
through the crossing. From the spectral properties discussed above we conclude
that $Z [\sigma ; \varepsilon]$ must be either symmetric or antisymmetric in
$\sigma$
\begin{equation}
  Z (\sigma ; \varepsilon) = \pi (\varepsilon) Z (- \sigma ; \varepsilon),
  \hspace{1em} \pi (\varepsilon) \in \{+ 1, - 1\}.
\end{equation}
In the next paragraph we shall show that
\begin{equation}
  \pi (\varepsilon) = \left\{ \begin{array}{lll}
    - 1 & \tmop{for} & \varepsilon_{\mu} = (0, 0)\\
    + 1 & \tmop{for} & \varepsilon_{\mu} = (1, 0), (0, 1), (1, 1)
  \end{array} \label{Pfarity} \right.
\end{equation}
holds.

We consider an interpolation
\begin{equation}
  g (t) = Z (t \sigma ; \varepsilon) = \tmop{Pf} [A (t \sigma)], \hspace{1em}
  t \in [- 1, 1]
\end{equation}
and assume $Z [\sigma ; \varepsilon] \not= 0$ as otherwise the proof is
trivial. Then the question of (anti)symmetry in $t$ can be `decided' in an
infinitesimal neighborhood of $t = 0$ which we can control by leading order
perturbation theory.

Except for $\varepsilon_{\mu} = (0, 0)$ the eigenvalues at $t = 0$ do not
vanish but are of order $L_{\mu}^{- 1}$. Hence there is no zero-crossing,
which proves the lower line of ($\ref{Pfarity}$). In the totally periodic case
there is a pair of $x$-independent zero-modes. The 2-spinors $u_{\pm}$ are
chosen to diagonalize the spin matrix of the perturbation $\mathcal{C}t
\sigma$
\begin{equation}
  \mathcal{C}u_{\pm} = \pm i u_{\pm}, \hspace{1em} A (0) u_{\pm} = 0.
\end{equation}
Now in leading order degenerate perturbation theory we find two eigenvalues
\begin{equation}
  \lambda_{\pm} \simeq \pm \frac{t}{L_0 L_1} \int d^2 x \sigma (x)
\end{equation}
associated with these eigenvectors.

Unless this integral vanishes we see that exactly one pair of eigenvalues
crosses zero as we move $t = 0 + \epsilon \rightarrow t = 0 - \epsilon$ and
hence the Pfaffian changes sign. The sign $\pi$ cannot jump with $\sigma$.
Thus, if the integral above vanishes, i.e. $\sigma$ has a vanishing
zero-momentum component, we deform $\sigma$ a little such that this is not the
case anymore. Now (\ref{Pfarity}) is established.

We note that our general result above is consistent with the explicit
evaluation in the continuum limit of the Wilson-Pfaffian for constant
$\sigma$, see appendix A in {\cite{Wolff:2007ip}}. Inspired by this
calculation we give another formal proof of the above result for a continuum
theory regularized by a cut-off in momentum space.
We start from
the identities
\begin{equation}
  \gamma_5^{\top} A (\sigma) \gamma_5 = A (- \sigma)
\end{equation}
and
\begin{equation}
  \tmop{Pf} [\gamma_5^{\top} A (\sigma) \gamma_5] = \tmop{Det} \gamma_5
  \tmop{Pf} [A (\sigma)] .
\end{equation}
The Det in the last line is taken in the function space over which
$\xi_{\alpha} (x)$ is integrated (with anticommuting expansion coefficients).
If Pf is written as a Grassmann integral, then $\tmop{Det} \gamma_5$ is the
Jacobian of $\xi \rightarrow \gamma_5 \xi$. Each $2 \times 2$ block
contributes a factor $\det \gamma_5 = - 1$ in the usual sense. We now imagine
to regularize the fermion integral by a finite mode number cutoff. If we
impose this cutoff in a way preserving the symmetry of rotations by $\pi$
then for each incorporated mode
$p_{\mu}$ also $- p_{\mu}$ has to be included. This leads to an odd total mode
number only for $\varepsilon_{\mu} = (0, 0)$ because of the unpaired mode
$p_{\mu} = (0, 0)$. Thus we have
\begin{equation}
  \tmop{Det} \gamma_5 = \pi (\varepsilon)
\end{equation}
representing a non-invariance of the integration measure in the case of fully
periodic boundary conditions. It also follows from the foregoing discussion
that expectation values in the GN model at odd $N$ with such boundary
conditions are ill defined at the chiral point in any finite volume, because
the normalizing partition function vanishes. \ This arises at least
superficially in a similar fashion as with the global anomaly in
{\cite{Witten:1982fp}}.

\section{Observables from fluctuating boundary conditions}

\subsection{Chiral order parameter}

We have introduced four different boundary conditions for the fermions in the
GN model. We may consider the choice among them as an additional dynamical
variable and consider general superpositions with an amplitude $z
(\varepsilon)$
\begin{equation}
  \tilde{Z} (\sigma ; z) = \sum_{\varepsilon} z (\varepsilon) Z (\sigma ;
  \varepsilon) . \label{zaver}
\end{equation}
Below we shall be interested in two special choices for the amplitudes
\begin{equation}
  z_+ (\varepsilon) = \frac{1}{2} - \delta_{\varepsilon, (0, 0)}, \hspace{1em}
  z_- (\varepsilon) = \frac{1}{2} . \label{zpm}
\end{equation}
The first choice is distinguished by leading to a strictly positive hopping or
world-line expansion for the finite volume GN model on the lattice
{\cite{Wolff:2007ip}}. The anomalous symmetry (including the sign of
Det$\gamma_5$) found in the previous section implies the relation
\begin{equation}
  \tilde{Z} (- \sigma ; z_+) = \tilde{Z} (\sigma ; z_-) .
\end{equation}
For either choice {\emdash} jointly for all flavors {\emdash} we find
partition functions for the interacting GN model
\begin{equation}
  Z^{\tmop{GN}}_{\pm} (m, g) = \int D \sigma \mathe^{- \frac{1}{2} \int d^2 x
  \sigma^2} [ \tilde{Z} (m + g \sigma ; z_{\pm})]^N \label{ZGN}
\end{equation}
which coincide at the chiral point $m = 0$ as one shows by transforming
$\sigma \leftrightarrow - \sigma$. We hence may define an order parameter that
vanishes at the chiral point
\begin{equation}
  \chi = \frac{1}{N} \ln (Z^{\tmop{GN}}_- / Z^{\tmop{GN}}_+) . \label{chidef}
\end{equation}

More conventional order parameters may be based on chirally odd observables \
$\overline{\xi} (x) \xi (y)$. In view of our later simulations we define
\begin{eqnarray}
  \langle \xi_{\alpha a} (u) \overline{\xi}_{\beta b} (v) \rangle_z & = &
  \text{$\frac{\delta_{a b}}{Z^{\tmop{GN}}_+}$}  \int D \sigma \mathe^{-
  \frac{1}{2} \int d^2 x \sigma^2} [ \tilde{Z} (m + g \sigma ; z)]^{N - 1} 
  \label{zcorr}\\
  & \times & \sum_{\varepsilon} z (\varepsilon) \int D_{\varepsilon} \xi
  \mathe^{- \frac{1}{2} \int d^2 x \overline{\xi} (\gamma_{\mu} \partial_{\mu}
  + m + g \sigma) \xi} \xi_{\alpha} (u) \overline{\xi}_{\beta} (v) . \nonumber
\end{eqnarray}
Here we indicated the periodicity in the integration measure. Note that we
always normalize with respect to the $z_+$ ensemble. Such correlations are $2
L_{\mu}$-periodic but neither periodic nor antiperiodic over the original
torus while bilinear local densities remain periodic. Chiral symmetry implies
\begin{equation}
  \gamma_5 \langle \xi_{\alpha a} (u) \overline{\xi}_{\beta b} (v)
  \rangle_{z_{\pm}} \gamma_5 = - \langle \xi_{\alpha a} (u)
  \overline{\xi}_{\beta b} (v) \rangle_{z_{\mp}} .
\end{equation}
An integration over a $2 L_1$ long interval projects to the spatially periodic
component at zero spatial momentum and we obtain the chirally odd scalar $k_S$
\begin{equation}
  k_S (x_0) = - \frac{1}{2 N} \int_{- L_1}^{+ L_1} d x_1  \left[ \langle
  \overline{\xi} (0) \xi (x) \rangle_{z_+} + \langle \overline{\xi} (0) \xi
  (x) \rangle_{z_-} \right] \label{kSdef}
\end{equation}
with both $\xi$ indices contracted here. For the purpose of monitoring the
field normalization we define in addition
\begin{equation}
  k_V (x_0) = - \frac{1}{N} \int_{- L_1}^{+ L_1} d x_1 \langle \overline{\xi}
  (0) \gamma_0 \xi (x) \rangle_{z_+} . \label{kVdef}
\end{equation}
We only use $z_+$ boundary conditions here, as they will lead to the highest
precision on the lattice. In this way $k_V$ has both chirally even and odd
components. Due to parity, $k_S$ is an even function of $x_0$ while $k_V$ is
odd.

We form the ratio $k_S / k_V$ to cancel multiplicative renormalization
factors. While with exact chiral symmetry $k_S$ vanishes, with a
regularization that violates chiral symmetry by effects linear in the lattice
spacing $a$ we can only expect to find $k_S (x_0) = \Omicron (a)$ if $\chi =
0$ has been imposed as a renormalization condition. We therefore tie $x_0$ to
a physical length scale to obtain a scaling situation and define
\begin{equation}
  \tilde{\chi}_n = \frac{k_S (n L_0 / 4)}{k_V (L_0 / 2)} \hspace{1em} (n = 1,
  2, 3)
\end{equation}
as universal quantities vanishing in the continuum limit at the chiral point.
The finite size $L_0$ supplies an infrared regulator and keeps our model
well-defined (for the boundary conditions chosen) at the chiral point in a
finite volume. This is similar to the Schr\"odinger functional of QCD
{\cite{Luscher:1992an}}, where the PCAC relation replaces the discrete chiral
symmetry relations invoked here.

\section{Lattice, Wilson fermions, loop gas}

To pass to the lattice formulation we restrict $\xi (x)$ to the sites of a
toroidal square lattice with spacing $a$ and integer
(anti)periodicity lengths $L_{\mu} / a$. In addition we replace the Dirac
operator in (\ref{Pf}) and ($\ref{Adef}$) by the Wilson matrix (with $r = 1$)
\begin{equation}
  \gamma_{\mu} \partial_{\mu} \rightarrow \gamma_{\mu} \tilde{\partial}_{\mu}
  - \frac{a}{2} \partial_{\mu} \partial_{\mu}^{\ast} .
\end{equation}
We use here standard notation and $\partial_{\mu}, \partial_{\mu}^{\ast},
\tilde{\partial}_{\mu}$ denote the forward, backward and symmetrized nearest
neighbor difference operators. The second term eliminates the doubler modes
but also breaks the chiral symmetry (\ref{chisym}) just like a mass term. The
consequences of chiral symmetry however are expected to emerge in the
continuum limit. To reach it one has to determine a critical value $m_c (g)$
as there is no symmetry to prevent an additive mass renormalization. From here
on we use lattice units and set $a = 1$.

\subsection{Loop gas representation of the GN model}

The Wilson-discretized GN model has already been studied in
{\cite{korzec:2006hy}}, {\cite{TomPhD}} by hybrid Monte Carlo methods and
Majorana-Wilson fermions were simulated in {\cite{Wolff:2007ip}} by a cluster
algorithm in the loop gas representation elaborating on
{\cite{Gattringer:1998cd}}. Most recently we have studied a worm algorithm
{\cite{Wolff:2008xa}} and we will now make extended use of results from this
paper which should be consulted for details.

In {\cite{Wolff:2007ip}} we have shown that the Grassmann integral
(\ref{zcorr}) {\emdash} for one flavor at this point {\emdash} can be written
as a sum over contributions from non-intersecting loops on the lattice with
one additional open string connecting the locations of the field insertions
$u$ and $v$. The underlying loop ensemble is
\begin{equation}
  \mathcal{Z}= \sum_{u, v, \{k (l)\}} \Theta (k ; u, v) 2^{- \overline{C} / 2}
  \prod_{x \in \mathcal{M}[k]} (2 + m + g \sigma (x)) . \label{loopens}
\end{equation}
In this formula the link variables $k (l) = 0, 1$ are one on the loops/string
and zero elsewhere. We also say that links with $k (l) = 1$ carry dimers. The
constraint $\Theta (k ; u, v) = 0, 1$ restricts the $k$ configurations to the
subset corresponding to the hopping expansion of the Majorana Wilson fermion
with field insertions at $u, v$. It demands that zero or two dimers are
attached to all sites except $u, v$ if they do not coincide. In this case they
must see exactly one dimer each, the beginning and end of the string
connecting them. The integer $\overline{C}$ counts corners, sites connected to
two orthogonal dimers. Finally the monomer set $\mathcal{M}[k]$ for each
configuration $k$ consists of all sites with no dimers attached.

For each $k$-configuration we define topological variables $e_{\mu} [k] \in
\{0, 1\}$ as follows. For each direction $\mu$ on the torus we distinguish one
sheet orthogonal to the $\mu$-direction (a line for $D = 2$) of links in that
direction. This is the same construction that is used in connection with the
center symmetry in lattice gauge theory. A definite convention is given in
{\cite{Wolff:2008xa}}. Then $e_{\mu}$ is zero or one depending on whether an
even or odd number of dimers belong to the sheet. Thus we decompose the loop
gas into four classes, see {\cite{Wolff:2007ip}}. If a contribution has $u =
v, \theta (k, u, u) = 1$ there are only closed loops and $e_{\mu}$ coincides
with their total mod 2 winding number in the respective direction. We also
associate possible antiperiodic boundary conditions with minus signs on bonds
in the respective sheets where we `close the torus'. Now the fermionic
partition function is given by
\begin{equation}
  \tilde{Z} (\sigma ; z) = \sum_k \Theta (k ; 0, 0) 2^{- \overline{C} / 2} 
  \prod_{x \in \mathcal{M}[k]} (2 + \sigma (x)) \Phi (e ; z)
\end{equation}
where we took $u = v = 0$ to only sum over closed loop `vacuum'
configurations. The amplitude $\Phi$ is
\begin{equation}
  \Phi (e ; z) = (2 \delta_{e, (0, 0)} - 1) \sum_{\varepsilon} z (\varepsilon)
  (- 1)^{\varepsilon \cdot e} .
\end{equation}
The first factor stems from spin and Fermi statistics in the Grassmann
integral (see {\cite{Wolff:2007ip}}, {\cite{Wolff:2008xa}}). In particular it
is +1 for all topologically trivial loop configurations which amounts to the
absence of a `hard' sign problem in two dimensions. The second factor
corresponds to the fluctuating boundary conditions. If we now insert $z =
z_{\pm}$ from (\ref{zpm}) we find
\begin{equation}
  \Phi (e ; z_+) = 1, \hspace{1em} \Phi (e ; z_-) = 2 \delta_{e, (0, 0)},
  \label{Phipm}
\end{equation}
and in particular the announced positivity for the $z_+$ ensemble.

To pass to the partition function of the interacting GN model we insert the
loop representation simultaneously for all flavors and find
\begin{equation}
  Z^{\tmop{GN}}_{\pm} (m, g) = \sum_{k_1 \ldots k_N}  \prod_{a = 1}^N [\Theta
  (k_a ; 0, 0) 2^{- \overline{C}_a / 2} \Phi (e_a ; z_{\pm})] \prod_x c (M
  (x)) .
\end{equation}
In this formula $k_a, \overline{C}_a, e_a$ are defined as before, just for
each flavor independently. The field $M (x)$ with possible values $0, 1,
\ldots, N$ depends on all $k_a$ (left implicit) and yields the total number of
monomers at $x$, i.e. it counts to how many $\mathcal{M}[k_a]$ the site $x$
belongs. The various flavors thus interact through the factor
\begin{equation}
  c (M) = \int_{- \infty}^{\infty} \frac{d \sigma}{\sqrt{2 \pi}} \mathe^{-
  \frac{1}{2} \sigma^2} [2 + m + g \sigma]^M . \label{cM}
\end{equation}

The weights $c (M)$ may conveniently be generated from a recursion formula. In
the Monte Carlo updates below we shall rather need ratios
\begin{equation}
  r (M) = \frac{c (M + 1)}{c (M) \text{}}
\end{equation}
which are directly generated by
\begin{equation}
  r (0) = 2 + m, \hspace{1em} r (M) = 2 + m + \frac{g^2 M}{r (M - 1)} .
\end{equation}

\subsection{GN worm algorithm}

Finally in our simulation we sample the ensemble
\begin{equation}
  \mathcal{Z}^{\tmop{GN}} = \sum_{j, u, v, \{k_a \}} \Theta (k_j ; u, v) 2^{-
  \overline{C}_j / 2} \prod_{a \not= j} [\Theta (k_a ; 0, 0) 2^{-
  \overline{C}_a / 2}] \prod_x c (M (x)) . \label{ZGNlG}
\end{equation}
The summation now includes a sum over $j = 1, 2, \ldots, N$ which is an
`active' flavor to which the insertions $u, v$ refer. A valid update procedure
is given by simply using the algorithm of {\cite{Wolff:2008xa}} for the active
flavor with the external field replaced $c (M (x))$ which depends also on the
non-active flavors that are momentarily frozen during the update. Whenever $u
= v$ is encountered the active flavor is changed randomly and ergodicity is
achieved. We summarize the whole procedure in the following paragraph.

The simulation alternates between analogous $u$-updates and $v$-updates, of
which we only detail the latter. In addition we alternate between type I and
II. Type I involves:
\begin{itemize}
  \item Pick a nearest neighbor $v'$ of $v$, call the link $l = \langle v v'
  \rangle$
  
  \item Propose the combined move $v \rightarrow v'$, $k_j (l) \rightarrow 1
  - k_j (l)$
  
  \item Compute the ratio $q$ of the weight given in {\cite{Wolff:2008xa}}
  after and before the move where we suitably substitute $r (M)$ for the
  external field ratio. The Metropolis decision is controlled by $|q|$.
\end{itemize}
In step II one only acts if $u = v$ is met. In this case we:
\begin{itemize}
  \item Pick a random new value $j \in \{1, 2, \ldots, N\}$ for the active
  flavor. We may actually imagine this as a Metropolis proposal tried after
  every type I move. Whenever $u \not= v$ holds it will be rejected due to
  the $\theta$-constraints while for $u = v$ it is accepted with probability
  one.
  
  \item With probability $p_{\tmop{kick}}$ we re-locate $u = v$ randomly on
  the lattice.
\end{itemize}

\subsection{Observables in the loop gas representation}

We define expectation values in the ensemble (\ref{ZGNlG})
\begin{equation}
  \langle \langle \mathcal{O} \rangle \rangle =
  \frac{1}{\mathcal{Z}^{\tmop{GN}}}  \sum_{j, u, v, \{k_a \}} \mathcal{O}
  \Theta (k_j ; u, v) 2^{- \overline{C}_j / 2} \prod_{a \not= j} [\Theta
  (k_a ; 0, 0) 2^{- \overline{C}_a / 2}] \prod_x c (M (x))
\end{equation}
where $\mathcal{O}$ may depend on all quantities that are summed over. The
chiral order parameter $\chi$ (\ref{chidef}) is easily measured by
\begin{equation}
  \mathe^{N \chi} = \frac{\langle \langle \prod_a \Phi (e_a ; z_-) \delta_{u,
  v} \rangle \rangle}{\langle \langle \delta_{u, v} \rangle \rangle} =
  \frac{\langle \langle \prod_a [2 \delta_{e_a, (0, 0)}] \delta_{u, v} \rangle
  \rangle}{\langle \langle \delta_{u, v} \rangle \rangle} . \label{chiobs}
\end{equation}
In other words, at the chiral point $\chi = 0$ the fraction $2^{- N}$ of all
vacuum ($u = v$) configurations has the trivial topology in all flavours
simultaneously. In the trivial case $g = 0$ this arises for $m = 0$ from
independent factors $1 / 2$ for each species.

Other observables considered in this paper are the correlations
\begin{equation}
  \langle \xi_a (x) \overline{\xi}_b (0) \rangle_z = \delta_{a b}
  \frac{\langle \langle \sum_{\varepsilon} z (\varepsilon) (- 1)^{\varepsilon
  \cdot e_j} \delta^{(\varepsilon)}_{x, u - v} \psi_j \mathcal{S}(k_j ; u, v)
  \prod_{c \not= j} \Phi (e_c ; z_{}) \rangle \rangle}{\langle \langle
  \delta_{u, v} \rangle \rangle} .
\end{equation}
Here $\delta^{(\varepsilon)}$ is the $\varepsilon$-(anti)periodic lattice
$\delta$ function and the spin matrix{\footnote{The case $u = v$ is special in
the definition of $\mathcal{S}$. It would depend on $M (u)$ here but will not
be required as we will not consider correlations with coinciding arguments ($x
= 0$).}} $\mathcal{S}$ has been defined in {\cite{Wolff:2008xa}}. It carries
two spin indices that we suppressed at $\xi, \overline{\xi}$. The sign
$\psi_j$ collects the phases from spin matrix elements and fermion loops in
the active flavor, see eq. (48) in {\cite{Wolff:2008xa}}. It is evolved during
the updates and thus available at every step. Applying this correspondence for
$k_S$ and $k_V$ in (\ref{kSdef}), (\ref{kVdef}), using (\ref{Phipm}), we
arrive at the translations
\begin{equation}
  k_V (x_0) = \frac{\langle \langle \psi_j \tmop{tr} [\gamma_0 \mathcal{S}(k_j
  ; u, v)] [(- 1)^{e_{0, j}} \delta^{(1)}_{x_0, u_0 - v_0} -
  \delta^{(0)}_{x_0, u_0 - v_0}] \rangle \rangle}{\langle \langle \delta_{u,
  v} \rangle \rangle}
\end{equation}
and
\begin{equation}
  k_S (x_0) = \frac{\langle \langle \psi_j \tmop{tr} [\mathcal{S}(k_j ; u, v)]
  [(- 1)^{e_{0, j}} \delta^{(1)}_{x_0, u_0 - v_0} E_j^+ - \delta^{(0)}_{x_0,
  u_0 - v_0} E_j^-] \rangle \rangle}{\langle \langle \delta_{u, v} \rangle
  \rangle}
\end{equation}
with
\begin{equation}
  E_j^{\pm} = \frac{1}{2} \left( 1 \pm \prod_{a \not= j} [2 \delta_{e_a, (0,
  0)}] \right) .
\end{equation}
At $x_0 = L_0 / 2$, due to parity covariance, $\delta^{(0)}$ in $k_V$ and
$\delta^{(1)}$ in $k_S$ do not contribute. Although these expressions may look
somewhat intimidating, all quantities entering are readily available from the
updates and the accumulation proceeds quickly in the form of histograms with
bins for each time separation $u_0 - v_0$ and each of the four possible $e_j$.
To debug the code we have found it useful to run at $g = 0$ comparing with the
known exact answers.

\section{Numerical tests in the Gross-Neveu model}

We have generalized the code described in {\cite{Wolff:2008xa}} to include an
arbitrary number of flavors coupled by the interaction term (\ref{cM}) and we
thus sample the ensemble (\ref{ZGNlG}). We have accumulated the necessary
observables entering into $\chi$ and $\tilde{\chi}_n, n = 1, 2, 3$. By
consulting perturbation theory and some preliminary experiments, we have run
series of simulations varying $L$ at fixed bare coupling $g$ with a value $m$
such that $\chi = 0$ holds up to a few sigma deviation. This is facilitated by
the fact that $m_c$ seems to hardly vary with $L$ at fixed $g$. For each of
the observables considered, for example those in (\ref{chiobs}), we also
compute its derivative with respect to $m$. This is achieved by introducing
another observable $W_m$, the derivative of the logarithm of the Boltzmann weight
with respect to $m$,
\begin{equation}
  \frac{\partial}{\partial m}  \prod_x c (M (x)) = W_m  \prod_x c (M (x)),
  \hspace{1em} W_m = \sum_x \frac{M (x)}{r (M (x) - 1)},
\end{equation}
which is also easy to keep track of during the simulations. Now
$m$-derivatives of not explicitly $m$-dependent observables $\mathcal{O}$ are
given by connected correlations
\begin{equation}
  \frac{\partial}{\partial m} \langle \langle \mathcal{O} \rangle \rangle =
  \langle \langle \mathcal{O}W_m \rangle \rangle - \langle \langle \mathcal{O}
  \rangle \rangle \langle \langle W_m \rangle \rangle .
\end{equation}
Thus we obtain estimates for $\partial \chi / \partial m$ and, forming
quotients, also $\partial \tilde{\chi}_n / \partial \chi$ (considering $g$
fixed). All these quantities are functions of primary expectation values and
they are estimated including errors by the technique presented in
{\cite{Wolff:2003sm}}. By exploiting the derivatives we can effectively
`post-run' vary $m$ a little bit to solve the tuning problem of finding the
value $m_c$ where $\chi$ vanishes. This is equivalent to reweighting in $m$
with a first order Taylor expansion of the Boltzmann factor. We thus for
example obtain an estimate for $m_c = m + \delta m$ with
\begin{equation}
  \delta m = - \chi \frac{1}{\partial \chi / \partial m} = m - \frac{\langle
  \langle \mathcal{O} \delta_{u, v} \rangle \rangle \langle \langle \delta_{u,
  v} \rangle \rangle \ln [\langle \langle \mathcal{O} \delta_{u, v} \rangle
  \rangle / \langle \langle \delta_{u, v} \rangle \rangle]}{\langle \langle
  \mathcal{O}W_m \delta_{u, v} \rangle \rangle \langle \langle \delta_{u, v}
  \rangle \rangle - \langle \langle \mathcal{O} \delta_{u, v} \rangle \rangle
  \langle \langle W_m \delta_{u, v} \rangle \rangle}
\end{equation}
with
\begin{equation}
  \mathcal{O}= \prod_a [2 \delta_{e_a, (0, 0)}]
\end{equation}
in this case. Since $\chi$ is expected to be a smooth function of $m L$ we
expect that derivatives with respect to $m L$ are order one. Then, for small
$\chi$ the next Taylor term can be roughly expected to be O($\chi \delta m$)
which we monitor to be negligible compared to the statistical error of $\delta
m$ and hence $m_c$. Where this not true, the initial tuning of $m$ was not
good enough for first-order reweighting and the run has to be repeated closer
to the chiral point. The range of possible corrections gets smaller as the
number of lattice sites grows because of the variance of the estimate of the
derivative.

The entries in table \ref{tab1} have been constructed in this way. Similarly
as in the above discussion we have estimated the combination $\tilde{\chi}_n -
( \text{$\partial \tilde{\chi}_n / \partial \chi$}) \chi$ to correct for tiny
nonzero $\chi$. Often this systematic correction is insignificant, within one
standard deviation, but the variance and thus the error is reduced due to
anticorrelated fluctuations of the terms. Each line in table \ref{tab1} is
based on a run of $8 \times 10^7$ iterations, where an iteration performs
$L^2$ elementary moves (counting $u$ and $v$).

\begin{table}[htb]
\centering
  \begin{tabular}{|l|l|l|l|l|l|l|l|}
    \hline
    $N$ & $g^2$ & $m$ & $L$ & $m_c$ & $\tilde{\chi}_1$ & $\tilde{\chi}_2$ &
    $\tilde{\chi}_3$\\
    \hline\hline
    2 & 1 & -0.4626 & 8 & -0.46249(22) & -0.0387(5) & -0.0252(4) &
    -0.0132(5)\\
    \hline
    2 & 1 & -0.4626 & 16 & -0.46236(15) & -0.0198(8) & -0.0134(6) &
    -0.0068(8)\\
    \hline
    2 & 1 & -0.4626 & 32 & -0.46254(13) & -0.0092(14) & -0.0074(9) &
    -0.0049(13)\\
    \hline
    2 & 1 & -0.4626 & 64 & -0.46275(14) & -0.0037(30) & -0.0030(20) &
    -0.0052(30)\\
    \hline\hline
    6 & 0.3 & -0.5967 & 8 & -0.59622(35) & -0.0379(19) & -0.0303(18) &
    -0.0236(17)\\
    \hline
    6 & 0.3 & -0.5967 & 16 & -0.59634(22) & -0.0263(27) & -0.0169(27) &
    -0.0116(26)\\
    \hline
    6 & 0.3 & -0.5967 & 32 & -0.59693(16) & -0.0182(44) & -0.0093(42) &
    -0.0099(40)\\
    \hline
    6 & 0.3 & -0.5967 & 64 & -0.59663(26) & -0.015(25) & -0.011(17) &
    -0.012(11)\\
    \hline\hline
    6 & 0.4 & -0.8007 & 8 & -0.80075(36) & -0.0558(19) & -0.0389(18) &
    -0.0286(17)\\
    \hline
    6 & 0.4 & -0.8007 & 16 & -0.80060(21) & -0.0312(29) & -0.0200(28) &
    -0.0121(26)\\
    \hline
    6 & 0.4 & -0.8007 & 32 & -0.80077(15) & -0.0246(47) & -0.0119(45) &
    -0.0014(42)\\
    \hline
    6 & 0.4 & -0.8007 & 64 & -0.80081(17) & -0.016(19) & -0.015(14) &
    -0.016(10)\\
    \hline\hline
    6 & 0.5 & -1.0058 & 8 & -1.00516(36) & -0.0707(19) & -0.0423(18) &
    -0.0256(18)\\
    \hline
    6 & 0.5 & -1.0058 & 16 & -1.00567(21) & -0.0471(31) & -0.0202(29) &
    -0.0071(28)\\
    \hline
    6 & 0.5 & -1.0058 & 32 & -1.00601(16) & -0.0250(56) & -0.0038(52) & 
    0.0066(50)\\
    \hline
    6 & 0.5 & -1.0063 & 64 & -1.00612(33) &  0.059(69) &  0.050(37) & 
    0.025(20)\\
    \hline
  \end{tabular}
  \caption{Simulation results for $m_c$ defined by $\chi = 0$ and variant
  chiral order parameters $\tilde{\chi}_n = \Omicron (a)$ at $m = m_c$.
  \label{tab1}}
\end{table}

\begin{figure}[htb]
  \begin{center}
    \epsfig{file=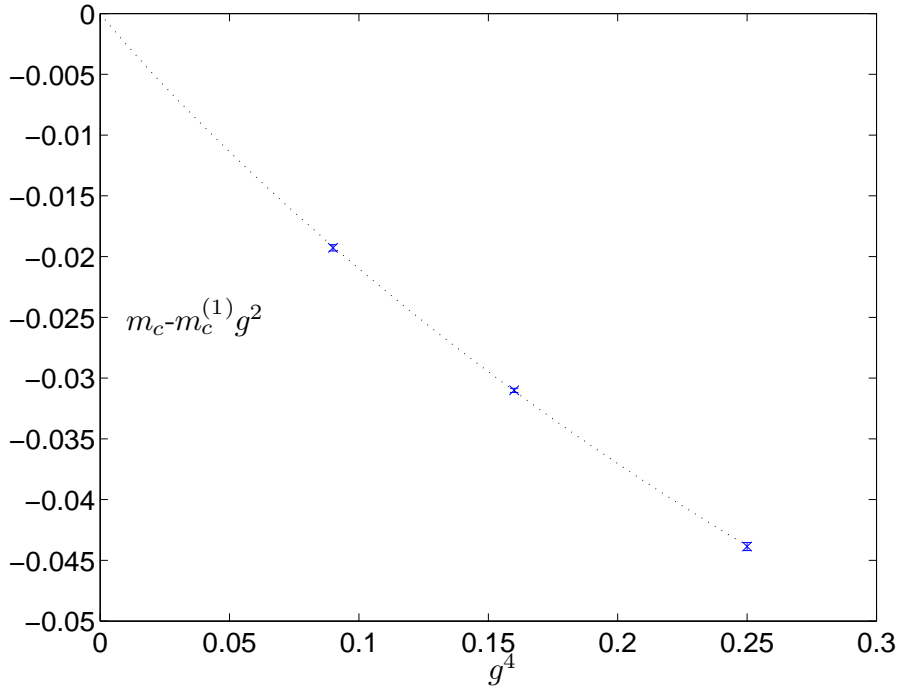,width=0.8\textwidth}
    \caption{Values of the critical mass $m_c$ in lattice units at $N = 6$
    after the subtraction of the 1-loop contribution.\label{fig1}}
  \end{center}
  \end{figure}

The first series is at $g^2 = 1, N = 2$ in the Thirring model. We expect no
divergent coupling renormalization in this case and a continuum limit could be
reached at fixed bare $g$, but the linearly divergent mass renormalization of
Wilson fermions is present. As an asymptotically free example, data were taken
for $N = 6$. We notice in the table that the $L$ dependence of $m_c$ is very
weak. A systematic trend is barely visible. In figure \ref{fig1} we plot our
results for $m_c$ (taken from $L = 64$) for $N = 6$, with the dominant one
loop contribution subtracted, against $g^4$. The dotted line corresponds to
the curve
\begin{equation}
  m_c \approx \sum_{n = 1}^3 \overline{m}_c^{(n)} g^{2 n}, \hspace{1em}
  \overline{m}_c^{(1)} = m_c^{(1)} \hspace{1em} \overline{m}_c^{(2)} = -
  0.270, \hspace{1em} \overline{m}_c^{(3)} = 0.189,
\end{equation}
where $m_c^{(1)}$ is the perturbative result (\ref{mc1}) (see appendix A).

{\noindent}It perfectly represents our data at the given error-level, but
$\overline{m}_c^{(i)}$ need not coincide with the true perturbative
coefficients $m_c^{(i)}, i = 2, 3$ that are not known to us.

For $N = 2$ the only physical scale is $L$ and we expect $\tilde{\chi}_n$ to
vanish proportional to $a / L$ as chiral symmetry emerges in the continuum
limit. A glance at the table shows that this is true within errors. For $N
\geqslant 3$ asymptotic freedom holds and the situation is different. The
continuum limit is expected to occur for $g^2 \searrow 0$. At least at large
$N$ {\cite{Gross:1974jv}} we know that the physics is governed by an effective
potential which for any $g^2$ has a critical $L_c$ such that there is one
minimum for $L < L_c$ but two minima for larger $L > L_c$
{\cite{Wolff:1985av}}. At $L = \infty$ a fermion mass $m_R$ is generated
spontaneously and emerges as a nonperturbative scale, somewhat analogous to
$\Lambda_{\tmop{QCD}}$. Therefore we expect for our $N = 6$ series a scaling
behavior
\begin{equation}
  \tilde{\chi}_n = \frac{a}{L} f_n (m_R L)
\end{equation}
with nontrivial but presumably smooth functions $f_n$.

At small $g$ we expect $m_R L$ to be tiny for $L \leqslant 64$ appearing in
the table and we may replace the scaling function by $f_n (m_R L) \approx f_n
(0)$ and thus the situation is similar to the Thirring model. This is still
roughly so for the couplings $g^2 \leqslant 0.5$ studied here. An
investigation of the GN model at large $N$ on the lattice with Wilson fermions
has suggested that for $N = 6, g^2 = 0.5$ the value $L_c$ may be not far from
our $L$ values, with quite some uncertainty however at finite $N$. On the
other hand the results in {\cite{Wolff:1985av}} suggest $m_R L_c = \Omicron
(1)$ and we may be close to the onset of a change in the scaling behavior.

In our data analysis we obtain information on autocorrelation times
$\tau_{\tmop{int}, \mathcal{O}}$ for all primary and derived observables that
we study. To avoid too much data handling we stored sub-histograms from 800
successive iterations resulting in a length of $10^5$ of our time series.
Between these blocked measurements autocorrelation times were small but
sometimes still measurable. Under these conditions our error estimates are
accurate at the percent level for the error of the error
{\cite{Wolff:2003sm}}. To obtain more detailed dynamical information we made
some shorter runs of $10^6$ blocks of only $N = 6$ iterations each. This has
yielded $\tau_{\tmop{int}, \chi} = \text{2.06(6), 3.47(14), 9.3(6), 69(10)}$
for $g^2 = 0.4$ and $L = 8, 16, 32, 64$. The unit of $\tau_{\tmop{int}}$ is
now `steps per site and flavor'. We see critical slowing down here, especially
for the step $32 \rightarrow 64$. This may be related to approaching a
physically `difficult' volume. The simulations of free fermions in
{\cite{Wolff:2008xa}} have shown very little slowing down. The less favorable,
but still not too bad, performance here is probably related to the flavors
holding back each other. With growing coupling the different flavors get more
more correlated, but they are only updated `one at a time' with the others
frozen until $u = v$ is met. One could in principle try to influence the
frequency of this by introducing a weight $\rho (u - v)$ as done in
{\cite{Wolff:2008km}} for the Ising model. In addition a second active flavor,
needed anyway to simulate four-point functions, could be beneficial here.

\section{Conclusions and outlook}

We have successfully incorporated the GN interaction in the loop gas
representation of Majorana fermions and found the algorithm of
{\cite{Wolff:2008xa}} to remain rather efficient although some critical
slowing down is visible. We have clarified the anomalous discrete chiral
symmetry behavior in a finite volume. It can be used to define the chiral
continuum limit with Wilson fermions by monitoring a topological observable
$\chi$ related to fluctuating boundary conditions in the finite volume loop
gas. It serves a purpose similar to the PCAC relation in the Schr\"odinger
functional scheme for QCD and makes the required mass fine-tuning quite
manageable. To complete the definition of a finite volume scheme for the GN
model we at present still lack a good definition of a running coupling in a
finite volume. The quantity proposed in {\cite{korzec:2006hy}} will presumably
be a good choice but it requires the extension of the algorithm to also sample
four-point functions. This seems feasible and will be required in any case to
be able to probe mass ratios for which exact conjectures have been made
{\cite{Dashen:1975xh}}. It also seems advisable to implement on-shell Symanzik
O($a)$ improvement as in QCD. The action has to be augmented by only one term,
$( \overline{\xi} \xi)^3$ for example, but in addition the elementary fermion
operator $\xi$ needs to be improved as it can mix with $\xi ( \overline{\xi}
\xi)$. It would presumably be a good idea, interesting in itself, to first
test this at large $N$ on the lattice with an analytic but nonperturbative
calculation. In the end it may hopefully also be possible to shed some light
on the special cases $N = 3, 4$ which may be supersymmetric models in disguise
{\cite{Witten:1977xv}}.

From the algorithmic point of view we have seen here that with the sign
problem absent in two dimensions (apart from a milder form in the finite size
effects) we could simulate a physically interesting interacting fermionic
theory with the all-order strong coupling/hopping reformulation as an
alternative to HMC methods {\cite{Duane:1987de}}.

\appendix\section{Some perturbative results}

In this appendix we explore in lattice perturbation theory to order $g^2$ some
of the quantities that were studied numerically before. To evaluate quantities
in leading order perturbation theory, it is useful to exploit the fact that
the Grassmann integrals for each individual flavor are linear in each $\sigma
(x)$ because it multiplies the bilinear $\overline{\xi} \xi$ at $x$ whose
square vanishes. Hence one derives
\begin{equation}
  \frac{\partial}{\partial \sigma (x)}  \tilde{Z} (\sigma ; z) = \tilde{Z}
  (\sigma ; z) \tilde{K} (\sigma ; z), \hspace{1em} \frac{\partial^2}{\partial
  \sigma (x)^2}  \tilde{Z} (\sigma ; z) = 0
\end{equation}
with
\begin{equation}
  \tilde{K} (\sigma ; z) = \frac{1}{2}  \text{tr$\tilde{G} (0 ; \sigma ; z$)}
\end{equation}
and
\begin{equation}
  \text{$\tilde{G} (x ; \sigma ; z$)=} \frac{\sum_{\varepsilon} z
  (\varepsilon) Z (\sigma ; \varepsilon) G (x ; \sigma ;
  \varepsilon)}{\tilde{Z} (\sigma ; z)} \label{Gtilde}
\end{equation}
which is given in terms of propagators $G (x ; \sigma ; \varepsilon)$ inverse
to $\gamma_{\mu} \tilde{\partial}_{\mu} - \frac{1}{2} \partial_{\mu}
\partial_{\mu}^{\ast} + \sigma$ with fixed boundary conditions $\varepsilon$.
Due to translation invariance we find
\begin{equation}
  \frac{\partial}{\partial \sigma (x)}  \tilde{Z} (m + g \sigma ; z) |_{g = 0}
  = \frac{g}{V}  \frac{\partial}{\partial m} \tilde{Z} (m ; z)
\end{equation}
with the volume $V = L_0 L_1$, and similarly
\begin{equation}
  \sum_x \frac{\partial}{\partial \sigma (x)}  \tilde{Z} (m + g \sigma ; z)
  \text{$\tilde{G} (x ; m + g \sigma ; z) |_{g = 0}$} = g
  \frac{\partial}{\partial m}  \tilde{Z} (m ; z) \text{$\tilde{G} (x ; m ;
  z$)} .
\end{equation}
In addition we can make use of
\begin{equation}
  \frac{\partial^2}{\partial \sigma (x)^2} \tilde{Z} (\sigma ; z)
  \text{$\tilde{G} (x ; \sigma ; z)$} = 0.
\end{equation}

There is a small subtlety here. The vanishing of the doubly periodic partition
function $Z (0, (0, 0)) = 0$ trivially implies $ \tilde{Z} (0 ; z_+) =
\tilde{Z} (0 ; z_-)$. A singularity in the propagator $G (x ; m ; (0, 0))
\simeq (m L_0 L_1)^{- 1} + \Omicron (m^0)$ lifts however the zero-mode and
yields an O($a$) difference
\begin{equation}
  \lim_{m \rightarrow 0} [ \tilde{K} (m ; z_-) - \tilde{K} (m ; z_+)] =
  \frac{2 c_{00}}{\sqrt{(L_0^2 + L_1^2) / 2}} . \label{delK}
\end{equation}
The constant derives from sums over momenta appropriate for the respective
boundary conditions and the numerical Symanzik expansion (see appendix D of
{\cite{Bode:1999sm}}) yields
\begin{equation}
  c_{00} = \frac{\lim_{m \rightarrow 0} Z (m ; (0, 0)) / (m
  L)}{\sum_{\varepsilon \not= (0, 0)} Z (0 ; \varepsilon)} = 0.155609865 -
  \frac{0.20668}{L^2} + \Omicron (L^{- 4}) .
\end{equation}
Here and below we perform the numerical evaluations for $L_0 = L_1 = L$. If we
write $\tilde{K} (0 ; z_-) - \tilde{K} (0 ; z_+)$ below we refer to
(\ref{delK}).

\subsection{Chiral order parameter $\chi$}

With the help of the above formulas it is now rather easy to derive
\begin{equation}
  Z^{\tmop{GN}}_{\pm} (m, g) = [ \tilde{Z} (m ; z_{\pm})]^N \left\{ 1 +
  \frac{g^2}{2} N (N - 1) V \tilde{K} (m ; z_{\pm})^2 \right\} + \Omicron
  (g^4)
\end{equation}
and hence
\begin{equation}
  \chi = \ln [ \tilde{Z} (m ; z_-) / \tilde{Z} (m ; z_+)] + \frac{g^2}{2} (N -
  1) V [ \tilde{K} (m ; z_-)^2 - \tilde{K} (m ; z_+)^2] + \Omicron (g^4)
\end{equation}
The perturbative Ansatz for the critical mass of Wilson fermions reads
\begin{equation}
  m_c (g^2) = g^2 m_c^{(1)} + g^4 m_c^{(2)} + \ldots .
\end{equation}
Here we may use to leading order
\begin{equation}
  Z^{\tmop{GN}}_{\pm} (g^2 m_c^{(1)}, 0) = [ \tilde{Z} (0 ; z_{\pm})]^N
  \left\{ 1 + g^2 m_c^{(1)} N V \tilde{K} (0 ; z_{\pm}) \right\} .
\end{equation}
With $\chi$ vanishing at tree level we now derive (up to O($g^4)$)
\begin{equation}
  \chi = \frac{V g^2}{2} [ \tilde{K} (0 ; z_-) - \tilde{K} (0 ; z_+)] \left\{
  (N - 1) [ \tilde{K} (0 ; z_-) + \tilde{K} (0 ; z_+)] + 2 m_c^{(1)} \right\}
  .
\end{equation}
The condition that $\chi$ vanishes as a chiral order parameter fixes the one
loop mass renormalization to
\begin{equation}
  m_c^{(1)} = - \frac{N - 1}{2} [ \tilde{K} (0 ; z_-) + \tilde{K} (0 ; z_+)] .
\end{equation}
Using the definitions this may also be written as
\begin{equation}
  m_c^{(1)} = - \frac{N - 1}{2}  \frac{\sum_{\varepsilon \not= (0, 0)} Z (0
  ; \varepsilon) \tmop{tr} G (0 ; 0 ; \varepsilon)}{\sum_{\varepsilon \not=
  (0, 0)} Z (0 ; \varepsilon)}
\end{equation}
and the numerical value is

\begin{equation}
  m_c^{(1)} = - (N - 1) 0.3849001795 - \frac{0.2439}{L^4} + \Omicron (L^{- 6})
  . \label{mc1}
\end{equation}

\subsection{Two-point correlator}

For the correlator (\ref{zcorr}) (for one flavor $a = b$) the following simple
result is obtained
\begin{equation}
  \langle \xi (x) \overline{\xi} (0) \rangle_z =
  \frac{Z^{\tmop{GN}}_z}{Z^{\tmop{GN}}_+}  \left\{ 1 + g^2 (N - 1) \tilde{K}
  (m ; z_{}) \frac{\partial}{\partial m} \right\} \text{$\tilde{G} (x ; m ;
  z$)} + \Omicron (g^4) .
\end{equation}
In this formula the partition function ratio can also be expanded and
contributes in general another O($g^2)$ term. At the chiral point however it
is unity for $z = z_-$ and trivially so for $z_+$. The correlations defined in
this paper now read
\begin{equation}
  k_V (x_0) = \left\{ 1 + g^2 (N - 1) \tilde{K} (m ; z_+)
  \frac{\partial}{\partial m} \right\} k^{(0)}_V (x_0) + \Omicron (g^4)
\end{equation}
and
\begin{equation}
  k_S (x_0) = \frac{1}{2} \sum_{\tau = +, -} \mathe^{N \chi \delta_{\tau, -}}
  \left\{ 1 + g^2 (N - 1) \tilde{K} (m ; z_{\tau}) \frac{\partial}{\partial m}
  \right\} k^{(0, \tau)}_S (x_0) + \Omicron (g^4) .
\end{equation}
Using results from appendix A in {\cite{Wolff:2008xa}} we find (for $0 < x_0 <
L_0$)
\begin{eqnarray}
  k^{(0)}_V (x_0) & = & \left. \frac{1}{(1 + m) \tilde{Z} (m ; z_+)} \times
  \right\{ Z (m, (1, 0)) \frac{\cosh [\omega (L_0 / 2 - x_0)]}{\cosh [\omega
  L_0 / 2]} \nonumber\\
  &  & \left. - Z (m, (0, 0)) \frac{\sinh [\omega (L_0 / 2 - x_0)]}{\sinh
  [\omega L_0 / 2]} \right\} 
\end{eqnarray}
where $\omega = \ln (1 + m)$ is the pole mass. For small $m$ we obtain
\begin{equation}
  (1 + m) k^{(0)}_V (x_0) \simeq 2 c_{10} (1 - c_{00} m L_0) - 2 c_{00} m (L_0
  - 2 x_0) + \Omicron (m^2)
\end{equation}
with another constant
\begin{equation}
  c_{10} = \frac{Z (0 ; (1, 0))}{\sum_{\varepsilon \not= (0, 0)} Z (0 ;
  \varepsilon)} = 0.313557560 - \frac{0.2199}{L^2} + \Omicron (L^{- 4}) .
\end{equation}
For the scalar correlator we derive
\begin{eqnarray}
  k^{(0, \pm)}_S (x_0) & = & \left. \frac{1}{(1 + m) \tilde{Z} (m ; z_{\pm})}
  \times \right\{ Z (m, (1, 0)) \frac{\sinh [\omega (L_0 / 2 - x_0)]}{\cosh
  [\omega L_0 / 2]} \nonumber\\
  &  & \left. \mp Z (m, (0, 0)) \frac{\cosh [\omega (L_0 / 2 - x_0)]}{\sinh
  [\omega L_0 / 2]} \right\} . 
\end{eqnarray}
In this case we find for small $m$
\begin{equation}
  (1 + m) k^{(0 \pm)}_S (x_0) = 2 c_{10} m (L_0 / 2 - x_0) \mp 4 c_{00} +
  \Omicron (m^2) .
\end{equation}
If we combine the above results we find that the alternative order parameters
$\tilde{\chi}_n$ vanish also at O($g^2)$ if we set $m = g^2 m_c^{(1)}$, i.e.
the O($a)$ contributions seen numerically start (presumably) at O($g^4)$ only.

{\noindent}{\bf Acknowledgments}. We would like to acknowledge helpful discussions
with Rainer Sommer. U.W. would like to thank Uwe-Jens Wiese and his colleagues at the university
of Bern for hospitality, where some part of this work was carried out.  Financial support
of the DFG via SFB transregio 9 is acknowledged.

\end{document}